\documentclass[12pt]{article}

\newif\ifanonymousdraft
\anonymousdraftfalse

\usepackage[a4paper,left=1in,right=1in,top=1.5in,bottom=1.5in]{geometry}
\usepackage[T1]{fontenc}
\usepackage[utf8]{inputenc}
\usepackage[british]{babel}
\usepackage{lmodern}
\usepackage{microtype}
\usepackage{setspace}
\usepackage{authblk}
\usepackage{amsmath,amssymb,amsthm,mathtools}
\usepackage{booktabs,array,longtable,tabularx}
\usepackage{graphicx}
\usepackage{float}
\graphicspath{{figures/}}
\usepackage{xcolor}
\usepackage{enumitem}
\usepackage{csquotes}
\usepackage{natbib}
\usepackage[colorlinks=true,allcolors=blue]{hyperref}
\usepackage[nameinlink,noabbrev]{cleveref}

\setlist[itemize]{leftmargin=1.5em,itemsep=0.2em,topsep=0.3em}

\newtheorem{assumption}{Assumption}
\newtheorem{lemma}{Lemma}
\newtheorem{proposition}{Proposition}
\newtheorem{corollary}{Corollary}
\theoremstyle{remark}

\DeclareMathOperator{\Var}{Var}
\DeclareMathOperator{\Cov}{Cov}
\newcommand{\R}{\mathbb{R}}
\newcommand{\ind}{\mathbf{1}}
\newcommand{\clamp}[1]{\left[#1\right]_{0}^{1}}

\newcommand{\E}{\mathbb{E}}
\newcommand{\phid}{\varphi}

\begin{document}
	
\ifanonymousdraft
\title{Fund Competition under Conflicting ESG Rating Methodologies}
\author{}
\else
\title{Fund Competition under Conflicting ESG Rating Methodologies\thanks{Helpful comments from Florian Berg, Enrico Onali, and Kevin Schneider are gratefully acknowledged. All errors and omissions remain the author's own.}}
\author{Wanling Rudkin\thanks{Corresponding author: University of Exeter Business School, Streatham Court, Rennes Drive, Exeter EX4 4PU, United Kingdom Email: \href{mailto:w.rudkin@exeter.ac.uk}{w.rudkin@exeter.ac.uk}; Tel: +44 (0)7514 023738}}
\affil{Finance and Accounting Department, University of Exeter}

\fi
\date{}	
\maketitle

\begin{abstract}
Competing ESG rating providers reward different portfolio attributes. This paper models funds that choose portfolios and fees for investors with heterogeneous ESG priorities. Portfolio changes can improve both providers’ scores or favour one methodology over the other, and investor demand determines which methodology each fund targets. Greater disagreement makes provider-specific positioning more productive but common improvement less productive. Funds therefore specialise more, yet both provider scores, investor participation, and equilibrium fees fall in the benchmark equilibrium. Investor heterogeneity creates matching gains from specialisation, while common improvement supports holdings-based ESG exposure. Methodology convergence improves participation and common exposure but weakens matching across investor clienteles. Convergence raises welfare when the social value of common exposure is sufficiently high.
\end{abstract}


\section{Introduction}
\label{sec:introduction}

ESG ratings are offered by ratings agencies who each have their own methodology. MSCI-IVA, Sustainalytics, and LSEG-ESG (formerly Refinitiv) apply competing methodologies to corporate ESG information and produce different score systems. MSCI-IVA and Sustainalytics both combine exposure and management, but MSCI-IVA reports managed exposure whereas Sustainalytics reports unmanaged risk; LSEG-ESG uses peer-relative percentile scores and a proprietary materiality matrix informed by disclosure \citep{quotb2026trading}. Methodological differences therefore reward different corporate and portfolio attributes. Funds are also provided by ratings from specialist agencies like Morningstar, as well as having ratings calculated from the ratings of the individual assets the fund holds. To improve their ESG ratings, fund managers must actively manage their portfolios. Morningstar sustainability rankings produced large fund-flow responses \citep{hartzmark2019investors}. Fund managers later altered holdings to improve sustainability ratings, although the rating-induced portfolio changes reduced financial performance once the return trade-off became salient \citep{gantchev2024sustainability}. Mechanical issuer-rating upgrades also increase selection by ESG funds even when underlying corporate activity does not change \citep{choi2026mechanical}.

Using MSCI-IVA, LSEG-ESG, and Sustainalytics, \citet{quotb2026trading} document that a one-standard-deviation increase in rating disagreement is associated with a 1.3\% decline in abnormal trading volume and wider bid--ask spreads. Stronger effects among norm-constrained institutional investors indicate that investor objectives shape the market response to conflicting ratings. Divergent ESG assessments therefore create two linked problems: uncertainty can reduce participation, while predictable methodology differences can redirect portfolio design. Scope, measurement, and weighting choices generate different evaluations of the same firm or portfolio \citep{berg2022aggregate}. A fund can improve one provider's score while lowering another provider's score, consuming organisational resources or weakening financial performance for investors.

Recognising the stylised facts of ESG rating disagreement, this paper develops a model of fund competition in which ratings divergence has an explicit portfolio geometry. Each fund selects a budget-neutral deviation from a financial benchmark along two orthogonal directions. Common effort moves the portfolio along the direction shared by both methodologies and raises both scores. Signed provider-specific effort moves the portfolio along the differential direction: positive effort raises Provider 1's score relative to Provider 2's, whereas negative effort raises Provider 2's score relative to Provider 1's. Both methodology vectors rotate symmetrically around an unchanged common direction, so greater disagreement changes score productivity without changing the portfolio attributes represented by the common and differential directions. Funds choose the sign of provider-specific effort rather than receiving a provider assignment by assumption.

Investor priorities determine the sign of provider-specific effort without requiring perfect sorting. Attached and contestable investors may have different distributions of methodology preferences over the ratings provided by the various rating agencies. Linear interior demand makes the conditional mean of each alignment distribution sufficient for fund demand. A fund tilts toward Provider 1 when its demand-weighted mean investor alignment is positive and toward Provider 2 when its demand-weighted mean investor alignment is negative. Opposite specialisation arises only when the two funds' effective alignments have opposite signs. A common provider preference can instead make both funds target the same methodology.

Portfolio geometry produces the paper's central comparative statics. Greater disagreement reduces the score loading on common effort and increases the loading on provider-specific effort. Equilibrium common effort falls, absolute specialisation expands, and cross-provider score spreads widen. Preference-consistent investor weights restrict provider-specific valuation to a component of total ESG valuation. Within partial coverage, the preference restriction makes disagreement lower even a fund's favoured systematic score, reduce participation, and compress fees despite stronger relative specialisation.

Cost incidence enters the model through two distinct channels. Baseline convex cost represents manager-level implementation resources, mandate constraints, and organisational effort paid by the fund owner rather than a return loss on every pound invested. A separate extension places a per-unit financial-performance loss in investor utility. Closed-form equilibrium survives: investor-borne losses attenuate common and provider-specific effort without changing the sign of provider targeting or the increase in relative specialisation.

A holdings-based outcome prevents reported agreement from being treated as truth. Outcome vector $a$ need not coincide with either methodology. Common portfolio effort improves holdings-based outcome $y$ only when the common portfolio direction loads positively on $a$. Equal and opposite tilts cancel in a symmetric market, while asymmetric fund sizes restore provider-specific outcome effects. Preference heterogeneity creates an additional matching gain because provider-specific portfolios serve investors with different priorities. Methodology convergence therefore changes product-market competition, investor matching, and the composition of capital allocated across portfolio attributes.

Three literatures frame the contribution of this paper. Sustainable-finance models study investor tastes, equilibrium returns, and rating uncertainty \citep{pastor2021sustainable,pedersen2021responsible,avramov2022sustainable,zerbib2022scapm}. ESG-rating models study information provision, business models, and incentives generated by multiple measures \citep{lovo2026who,chaigneau2025executive}. Mutual-fund models study search, differentiation, portfolio overlap, and fees \citep{hortacsu2004product,wahal2011competition,li2014financial}. The contribution differs from all three literatures. Conflicting methodologies change the production technology of financial products, segment fund demand according to investor priorities, and alter fee competition through endogenous portfolio design.

Section~\ref{sec:institutional} develops methodology geometry, portfolio outcomes, and investor demand. Section~\ref{sec:equilibrium} solves positioning across coverage regimes and endogenises fees. Section~\ref{sec:ratingsystems} extends the technology to several providers and derives endogenous reliance. Section~\ref{sec:welfare} studies welfare and methodology convergence. Section~\ref{sec:implications} presents numerical and empirical implications. Section~\ref{sec:conclusion} concludes.

\section{Institutional setting and model}
\label{sec:institutional}

Fund ratings aggregate issuer-level ESG information into portfolio-level assessments. MSCI-IVA reports managed exposure on a higher-is-better scale, whereas Sustainalytics reports unmanaged risk on a higher-is-worse scale. LSEG-ESG uses peer-relative percentile rankings and a proprietary materiality matrix informed by disclosure \citep{quotb2026trading}. Coverage rules, category definitions, missing-data treatments, materiality assessments, score directions, and aggregation weights therefore differ across providers. Berg, K\"olbel, and Rigobon attribute rating divergence to measurement, scope, and weighting choices and document a provider-specific rater effect \citep{berg2022aggregate}. The model developed here separates systematic methodology geometry from measurement uncertainty.

Recent trading evidence adds a participation channel to the portfolio-design mechanism. \citet{quotb2026trading} associate rating disagreement with lower abnormal trading volume and wider bid--ask spreads, with stronger effects among norm-constrained institutional investors. Lower participation captures withdrawal under uncertainty, while heterogeneous methodology alignment captures differences across investor objectives. Sustainability labels and rankings also move mutual-fund flows \citep{hartzmark2019investors,ceccarelli2024low}. Managers alter holdings after rating incentives become salient, and the resulting trades can sacrifice financial performance \citep{gantchev2024sustainability}. Mechanical rating changes affect ESG-fund security selection even when underlying corporate activity remains unchanged \citep{choi2026mechanical}. Rating-sensitive demand therefore creates portfolio-design incentives, while organisational limits and investor-borne financial losses constrain portfolio adjustment.

Fund competition combines persistent clienteles with active comparison. Search costs, distribution arrangements, brand familiarity, and retirement-plan menus create local demand \citep{hortacsu2004product}. Portfolio overlap intensifies competition and induces incumbent funds to reduce management fees \citep{wahal2011competition}. Differentiation across factor positions creates market power and supports higher fees \citep{li2014financial}. Discrete Morningstar rating changes also influence flows independently of underlying performance measures \citep{delguercio2008star}. Evidence on search, portfolio overlap, differentiation, and rating-sensitive flows motivates joint portfolio and fee choices within the model.

\subsection{Funds, portfolios, and methodology geometry}
\label{sec:portfolio}

A strictly interior benchmark portfolio $w_0\in\mathbb R^K$ satisfies $\mathbf 1'w_0=1$ and $w_{0k}>0$ for every asset. Matrix $N\in\mathbb R^{K\times(K-1)}$ spans the budget-neutral tangent space, with full column rank and $\mathbf 1'N=0$. Fund portfolios take the form
\begin{equation}
w=w_0+Nz,
\label{eq:budgetportfolio}
\end{equation}
where reduced coordinate $z\in\mathbb R^{K-1}$ automatically preserves the budget constraint.

Positive-definite matrix $H$ measures manager-level difficulty in implementing a benchmark deviation. Monetary cost
\begin{equation}
\mathcal K(z)=\frac{\kappa}{2}z'Hz
\label{eq:financialloss}
\end{equation}
is paid by the fund owner and does not scale mechanically with assets under management. Organisational research, mandate governance, benchmark constraints, and fixed implementation capacity provide natural interpretations. Section~\ref{sec:costincidence} separately introduces financial-performance losses borne per invested account.

Provider $r$ assigns portfolio rating $S_r(w)$. A methodology cell is a region of portfolio space over which a provider applies fixed scoring weights. Core analysis applies inside a methodology cell on which the rating map is affine:
\begin{equation}
S_r(w_0+Nz)-S_r(w_0)=b_r'z,
\label{eq:localrating}
\end{equation}
where $b_r=N'\nabla_wS_r(w_0)$ is the reduced methodology gradient. Piecewise-linear scoring rules satisfy affine equation~\eqref{eq:localrating} between category or coverage thresholds. Appendix~\ref{sec:nonlinearrobustness} provides a perturbation bound for smooth nonlinear ratings.

A separate holdings-based ESG outcome is
\begin{equation}
y=a'z,
\label{eq:outcome}
\end{equation}
where reduced vector $a$ maps budget-neutral portfolio deviations into the environmental or social exposure relevant for welfare. Neither provider need use $a$ as its methodology.

Let $B$ collect methodology vectors as rows and define
\begin{equation}
G=BH^{-1}B'.
\label{eq:gram}
\end{equation}
Matrix $G$ is the methodology Gram matrix under organisational metric $H^{-1}$.

\begin{lemma}[Portfolio implementation]
\label{lem:implementation}
Suppose $B$ has full row rank. Minimum-cost reduced deviation attaining systematic score vector $s$ is
\begin{equation}
z^*=H^{-1}B'(BH^{-1}B')^{-1}s,
\label{eq:minportfolio}
\end{equation}
and its cost is
\begin{equation}
C(s)=\frac{\kappa}{2}s'G^{-1}s.
\label{eq:generalcost}
\end{equation}
Holdings-based ESG outcome at the minimum-cost portfolio $z^*$ is
\begin{equation}
y(s)=a'H^{-1}B'G^{-1}s.
\label{eq:outcomemap}
\end{equation}
\end{lemma}

Two-provider analysis holds the methodology bisector fixed. Common direction $u$ and differential direction $v$ satisfy
\begin{equation}
u'H^{-1}u=v'H^{-1}v=1,
\qquad
u'H^{-1}v=0.
\label{eq:uvorthogonal}
\end{equation}
For disagreement $\Delta\in(0,2)$, define
\begin{equation}
\gamma_\Delta=\sqrt{1-\frac{\Delta}{2}},
\qquad
\delta_\Delta=\sqrt{\frac{\Delta}{2}},
\label{eq:gammadelta}
\end{equation}
and methodology gradients
\begin{equation}
b_1(\Delta)=\gamma_\Delta u+\delta_\Delta v,
\qquad
b_2(\Delta)=\gamma_\Delta u-\delta_\Delta v.
\label{eq:methodologyfamily}
\end{equation}
Both gradients have unit norm under $H^{-1}$, while
\begin{equation}
b_1'H^{-1}b_2=1-\Delta.
\label{eq:similarity}
\end{equation}
Parameter $\Delta$ spans positive similarity, orthogonality, and negative similarity. Increasing $\Delta$ rotates the two methodologies symmetrically away from an invariant common direction $u$ toward opposing loadings on $v$. Section~\ref{sec:convergence} states how outcome conclusions change when the bisector itself rotates.

Minimum-cost reduced deviations in the span of $u$ and $v$ can be written
\begin{equation}
z_j=H^{-1}(x_j u+r_j v).
\label{eq:effortportfolio}
\end{equation}
Common portfolio effort is $x_j$. Signed provider-specific effort is $r_j$: positive values favour Provider 1 and negative values favour Provider 2. Systematic scores become
\begin{equation}
s_{1j}=\gamma_\Delta x_j+\delta_\Delta r_j,
\qquad
s_{2j}=\gamma_\Delta x_j-\delta_\Delta r_j.
\label{eq:scoresxr}
\end{equation}
Define common score component and signed score tilt as
\begin{equation}
g_j=\gamma_\Delta x_j,
\qquad
d_j=\delta_\Delta r_j.
\label{eq:gddef}
\end{equation}
Manager-level implementation cost and holdings-based outcome reduce to
\begin{equation}
C_j=\frac{\kappa}{2}(x_j^2+r_j^2),
\qquad
y_j=\theta_u x_j+\theta_v r_j,
\label{eq:costoutcomexr}
\end{equation}
where
\begin{equation}
\theta_u=a'H^{-1}u,
\qquad
\theta_v=a'H^{-1}v.
\label{eq:thetaloadings}
\end{equation}
Equivalent score-space cost is
\begin{equation}
C_\Delta(g,d)=\kappa\left(\frac{g^2}{2-\Delta}+\frac{d^2}{\Delta}\right).
\label{eq:scorecost}
\end{equation}

\begin{assumption}[Common-direction informativeness]
\label{ass:theta}
Holdings-based outcome loading satisfies $\theta_u\geq0$.
\end{assumption}

Positive $\theta_u$ means common portfolio effort improves holdings-based outcome $y$. No sign restriction is imposed on $\theta_v$ because provider-specific portfolio priorities can align positively or negatively with outcome vector $a$.

\begin{assumption}[Feasible affine cell]
\label{ass:trust}
A radius $R>0$ exists such that every equilibrium and admissible deviation satisfies
\begin{equation}
z'Hz=x_j^2+r_j^2<R^2
\label{eq:trustregion}
\end{equation}
and $w_0+Nz\geq0$. Rating equation~\eqref{eq:localrating} holds exactly throughout the feasible affine cell.
\end{assumption}

Strict interiority of $w_0$ guarantees that a sufficiently small $R$ preserves long-only feasibility. Assumption~\ref{ass:trust} therefore imposes budget balance, non-negative portfolio weights, and an exact affine rating technology within one economically relevant neighbourhood.

\subsection{Investors, participation, and fund demand}
\label{sec:demand}

Investor ESG priorities provide a foundation for provider alignment. Investor $i$ places relative weight $p_i\in[0,1]$ on attributes emphasised by Provider 1 and $1-p_i$ on attributes emphasised by Provider 2. Valuation of systematic scores is
\begin{equation}
p_i s_{1j}+(1-p_i)s_{2j}
=\gamma_\Delta x_j+\nu_i\delta_\Delta r_j,
\qquad
\nu_i=2p_i-1\in[-1,1].
\label{eq:investoraggregation}
\end{equation}
Willingness to pay for ESG attributes is scaled by $\alpha>0$.

A fraction $1-\lambda$ belongs to attached fund segments split equally between the funds. Search costs, distribution arrangements, brand familiarity, and retirement-plan menus generate attachment. Let $F_j(\nu)$ denote the alignment distribution in Fund $j$'s attached segment and define
\begin{equation}
\nu_j=\int_{-1}^{1}\nu\,dF_j(\nu).
\label{eq:clientelemean}
\end{equation}
No perfect sorting is imposed: $F_j$ may contain investors aligned with either provider. A fraction $\lambda$ of investors is contestable, and contestable alignment follows distribution $F_C(\nu)$ with mean
\begin{equation}
\nu_C=\int_{-1}^{1}\nu\,dF_C(\nu).
\label{eq:contestmean}
\end{equation}
Baseline symmetry later uses $\nu_A=\nu>0$, $\nu_B=-\nu$, and $\nu_C=0$, but all provider-direction results first allow arbitrary means.

A local investor attached to Fund $j$ has non-ESG mismatch $\xi_i\sim U[0,1]$ and receives
\begin{equation}
U_{ij}^{L}
=v-f_j+\alpha\gamma_\Delta x_j
+\alpha\nu_i\delta_\Delta r_j-t\xi_i.
\label{eq:localutility}
\end{equation}
Outside option utility equals zero. Contestable investors always participate and draw relative fund taste $\varepsilon\sim U[-h,h]$, independently of $\nu_i$. Absolute utilities are
\begin{align}
U_A^C&=\bar v-f_A+\alpha\gamma_\Delta x_A
+\alpha\nu_i\delta_\Delta r_A+\frac{\varepsilon}{2},
\label{eq:contestA}\\
U_B^C&=\bar v-f_B+\alpha\gamma_\Delta x_B
+\alpha\nu_i\delta_\Delta r_B-\frac{\varepsilon}{2},
\label{eq:contestB}
\end{align}
Assume $\bar v$ is large enough that every contestable investor prefers one of the two funds to the outside option.

Define
\begin{equation}
\ell=\frac{1-\lambda}{2t},
\qquad
c=\frac{\lambda}{2h},
\qquad
b=\ell+c.
\label{eq:lcb}
\end{equation}
Local participation for alignment type $\nu$ is
\begin{equation}
p_j(\nu)=\clamp{\frac{v-f_j+\alpha\gamma_\Delta x_j+\alpha\nu\delta_\Delta r_j}{t}}.
\label{eq:localparticipation}
\end{equation}
Interior contestable aggregation depends only on $\nu_C$:
\begin{equation}
S_A=\clamp{\frac12+
\frac{\alpha\gamma_\Delta(x_A-x_B)
+\alpha\nu_C\delta_\Delta(r_A-r_B)
-(f_A-f_B)}{2h}}.
\label{eq:contestshare}
\end{equation}
Total demand is
\begin{equation}
D_A=\frac{1-\lambda}{2}\int p_A(\nu)\,dF_A(\nu)+\lambda S_A,
\label{eq:demandA}
\end{equation}
with a symmetric expression for Fund $B$.

\begin{assumption}[Support-wide interior choice]
\label{ass:contest}
For every portfolio-fee profile considered in an interior equilibrium, each contestable type satisfies
\begin{equation}
\left|\alpha\gamma_\Delta(x_A-x_B)
+\alpha\nu\delta_\Delta(r_A-r_B)
-(f_A-f_B)\right|<h
\quad
\text{for all }\nu\in\operatorname{supp}(F_C).
\label{eq:contestinterior}
\end{equation}
Whenever partial local coverage is invoked, each attached type also satisfies
\begin{equation}
0<v-f_j+\alpha\gamma_\Delta x_j
+\alpha\nu\delta_\Delta r_j<t
\quad
\text{for all }\nu\in\operatorname{supp}(F_j).
\label{eq:localtypeinterior}
\end{equation}
\end{assumption}

Support-wide interiority makes type-level demand linear before expectations are taken. Contestable demand therefore reduces exactly to equation~\eqref{eq:contestshare}, and attached demand depends on alignment distribution $F_j$ only through mean $\nu_j$. Type-level saturation would make higher moments relevant for demand as well as welfare.

\subsection{Timing}
\label{sec:timing}

Funds choose portfolio efforts first and fees second. Reported ratings are then published and investors allocate capital. Baseline investors allocate capital using deterministic systematic scores. Appendix~\ref{sec:measurement} introduces expected demand under noisy publication.

Sequential directional-location timing, as applied in \cite{lai2001sequential},  provides limiting intuition rather than a separate model block amongst the generalised approach in this paper. Departure from \cite{lai2001sequential}'s benchmark arises because funds choose portfolios in several directions, incur manager-level implementation costs, compete for contestable investors, and set fees.

\section{Equilibrium with conflicting methodologies}
\label{sec:equilibrium}

Fixed-margin analysis first isolates portfolio positioning and market coverage. Endogenous fee setting then introduces price competition. Signed provider-specific effort allows each fund to choose which methodology to favour.

\subsection{Signed provider direction and coverage regimes}
\label{sec:coverage}

A fixed margin $m>0$ per investor gives Fund $j$ profit
\begin{equation}
\pi_j=mD_j-\frac{\kappa}{2}(x_j^2+r_j^2).
\label{eq:fixedprofit}
\end{equation}
Manager-level implementation cost is paid by the fund owner. Attention is restricted to parameter values for which every attached type obtains positive gross utility. Partial coverage requires gross utility below $t$ for every attached type; full coverage requires gross utility of at least $t$ for every attached type.

Define Fund $j$'s effective alignment as the sum of attached and contestable mean alignments weighted by their marginal demand coefficients:
\begin{equation}
\zeta_j=\ell\nu_j+c\nu_C.
\label{eq:effectivealignment}
\end{equation}

\begin{proposition}[Endogenous provider direction]
\label{prop:direction}
Suppose every attached alignment type has partial coverage and Assumptions~\ref{ass:trust}--\ref{ass:contest} hold. Each fund's unique interior portfolio choice is
\begin{equation}
x_j^P=\frac{m\alpha b\gamma_\Delta}{\kappa},
\qquad
r_j^P=\frac{m\alpha\zeta_j\delta_\Delta}{\kappa}.
\label{eq:generalpartial}
\end{equation}
Consequently,
\begin{equation}
\operatorname{sign}(r_j^P)=\operatorname{sign}(\zeta_j).
\label{eq:signdirection}
\end{equation}
Funds target opposite providers if and only if $\zeta_A\zeta_B<0$.
\end{proposition}

Provider targeting follows the demand-weighted mean alignment of attached and contestable investors. Opposite targeting need not be assumed and need not occur. A common contestable preference can draw both funds toward the same provider even when attached segments differ.

Coverage-regime derivations further specialise to homogeneous attached alignments, with $F_A$ concentrated at $\nu>0$, $F_B$ concentrated at $-\nu$, and $\nu_C=0$. Proposition~\ref{prop:direction} remains valid for arbitrary alignment distributions under partial coverage. Let $r\geq0$ denote Fund $A$'s effort magnitude, with Fund $B$ choosing $-r$. Representative local gross utility before fees is
\begin{equation}
q(x,r)=v+\alpha\gamma_\Delta x+\alpha\nu\delta_\Delta r.
\label{eq:qxr}
\end{equation}
Proposition~\ref{prop:direction} yields the partial-coverage candidate
\begin{equation}
x_P=\frac{m\alpha b\gamma_\Delta}{\kappa},
\qquad
r_P=\frac{m\alpha\nu\ell\delta_\Delta}{\kappa},
\label{eq:partialcandidate}
\end{equation}
with $q_P=q(x_P,r_P)$.

Once every attached investor participates, additional portfolio utility no longer expands attached demand. Common effort still reallocates contestable investors, yielding
\begin{equation}
x_F=\frac{m\alpha c\gamma_\Delta}{\kappa},
\qquad
r_F=0,
\label{eq:fullcandidate}
\end{equation}
with $q_F=v+\alpha\gamma_\Delta x_F$.

When $q_F<t\leq q_P$, equilibrium lies at the coverage boundary. Define $R_t=t-v$, $a_x=\alpha\gamma_\Delta$, and $a_r=\alpha\nu\delta_\Delta$. Boundary effort is
\begin{align}
x_B&=x_F+\frac{a_x(R_t-a_xx_F)}{a_x^2+a_r^2},
\label{eq:boundaryx}\\
r_B&=\frac{a_r(R_t-a_xx_F)}{a_x^2+a_r^2}.
\label{eq:boundaryr}
\end{align}
Boundary choices $x_B$ and $r_B$ satisfy $a_xx_B+a_rr_B=R_t$.

\begin{proposition}[Coverage regimes]
\label{prop:regimes}
Maintain positive local participation, Assumptions~\ref{ass:trust}--\ref{ass:contest}, and strict trust-region validity.
\begin{enumerate}[label=(\roman*)]
\item If $q_P<t$, a unique symmetric equilibrium within the maintained local model has partial coverage and equals $(x_P,r_P)$.
\item If $q_F\geq t$, a unique symmetric equilibrium within the maintained local model has full coverage and equals $(x_F,0)$.
\item If $q_F<t\leq q_P$, a unique symmetric equilibrium within the maintained local model lies at the coverage boundary and equals $(x_B,r_B)$.
\end{enumerate}
\end{proposition}

Market coverage determines whether portfolio attributes attract additional investors or only reallocate contestable investors. Partial-coverage funds use common and provider-specific effort. Full-coverage funds stop targeting provider-specific preferences because every local investor already participates. Boundary funds provide exactly enough investor value to cover their attached segments.

Common and provider-specific score components in the partial regime are
\begin{equation}
g_P=\gamma_\Delta x_P
=\frac{m\alpha b(2-\Delta)}{2\kappa},
\qquad
d_P=\delta_\Delta r_P
=\frac{m\alpha\nu\ell\Delta}{2\kappa}.
\label{eq:gdpartial}
\end{equation}
Fund $A$ reports $(g_P+d_P,g_P-d_P)$; Fund $B$ reports the reverse pair.

\begin{proposition}[Disagreement and specialisation]
\label{prop:reallocation}
Suppose the symmetric equilibrium has partial coverage. Greater disagreement satisfies
\begin{align}
\frac{\partial x_P}{\partial\Delta}
&=-\frac{m\alpha b}{4\kappa\gamma_\Delta}<0,
&
\frac{\partial r_P}{\partial\Delta}
&=\frac{m\alpha\nu\ell}{4\kappa\delta_\Delta}>0,
\label{eq:xrderivatives}\\
\frac{\partial g_P}{\partial\Delta}
&=-\frac{m\alpha b}{2\kappa}<0,
&
\frac{\partial d_P}{\partial\Delta}
&=\frac{m\alpha\nu\ell}{2\kappa}>0.
\label{eq:gdderivatives}
\end{align}
Whenever $c>0$ or $\nu<1$, both a fund's favoured score and its unfavoured score fall:
\begin{align}
\frac{\partial(g_P+d_P)}{\partial\Delta}
&=-\frac{m\alpha}{2\kappa}\left[c+\ell(1-\nu)\right]<0,
\label{eq:favouredfall}\\
\frac{\partial(g_P-d_P)}{\partial\Delta}
&=-\frac{m\alpha}{2\kappa}\left[b+\nu\ell\right]<0.
\label{eq:unfavouredfall}
\end{align}
Total participation also falls because
\begin{equation}
\frac{\partial q_P}{\partial\Delta}
=-\frac{m\alpha^2}{2\kappa}
\left[c+\ell(1-\nu^2)\right]<0.
\label{eq:qfall}
\end{equation}
\end{proposition}

Preference-consistent methodology weights produce a sharper conclusion than an unrestricted tilt coefficient. Disagreement raises relative specialisation but does not raise a fund's favoured score. Provider-specific valuation remains bounded by total ESG valuation, so cheaper provider-specific effort cannot offset the loss of common score productivity when contestable investors are present.

With equal fund demands, average holdings-based outcome is
\begin{equation}
\bar y_P=\theta_u x_P.
\label{eq:averageoutcome}
\end{equation}
Under Assumption~\ref{ass:theta}, disagreement weakly lowers average holdings-based outcome and lowers average outcome strictly when $\theta_u>0$.

\subsection{Endogenous fees}
\label{sec:fees}

Funds now choose portfolio efforts in stage one and fees in stage two. Fee-setting analysis maintains the symmetric benchmark with $\nu_A=\nu$, $\nu_B=-\nu$, and $\nu_C=0$, together with partial local coverage and interior contestable demand. Conditional on effort choices, Fund $A$ demand can be written
\begin{equation}
D_A=A_A-bf_A+cf_B,
\label{eq:feedemand}
\end{equation}
where
\begin{equation}
A_A=\ell\left(v+\alpha\gamma_\Delta x_A
+\alpha\nu_A\delta_\Delta r_A\right)
+\frac{\lambda}{2}
+c\left[\alpha\gamma_\Delta(x_A-x_B)
+\alpha\nu_C\delta_\Delta(r_A-r_B)\right].
\label{eq:AA}
\end{equation}
Quantity $A_A$ is Fund $A$'s fee-independent demand intercept. Define
\begin{equation}
M=4b^2-c^2,
\qquad
Z=2b-c=2\ell+c.
\label{eq:MZ}
\end{equation}
Quantity $M$ is the determinant of the fee first-order-condition system, while $Z$ is the coefficient governing a symmetric fee change.

\begin{lemma}[Fee subgame]
\label{lem:fee}
For effort choices satisfying the interior-demand conditions, the fee subgame has the unique equilibrium
\begin{align}
f_A&=\frac{2bA_A+cA_B}{M},
\label{eq:feeA}\\
f_B&=\frac{cA_A+2bA_B}{M}.
\label{eq:feeB}
\end{align}
At a symmetric effort allocation,
\begin{equation}
f=\frac{\ell(v+\alpha\gamma_\Delta x+\alpha\nu\delta_\Delta r)+\lambda/2}{Z},
\qquad
D=bf.
\label{eq:symmetricfee}
\end{equation}
\end{lemma}

Fee pass-through coefficients measure how a fund's own common and provider-specific effort affect its equilibrium fee. Define
\begin{equation}
\phi_g=\frac{2b^2-c^2}{M},
\qquad
\phi_d=\frac{2b\ell}{M},
\label{eq:rgrd}
\end{equation}
and effort multipliers
\begin{equation}
\mathcal X(\Delta)=\frac{2b\alpha \phi_g\gamma_\Delta}{\kappa},
\qquad
\mathcal R(\Delta)=\frac{2b\alpha\nu \phi_d\delta_\Delta}{\kappa}.
\label{eq:XR}
\end{equation}
Stage-one strict concavity requires
\begin{equation}
\Psi(\Delta)=\frac{2b\alpha^2}{\kappa}
\left(\gamma_\Delta^2\phi_g^2+\nu^2\delta_\Delta^2\phi_d^2\right)<1.
\label{eq:concavity}
\end{equation}

\begin{proposition}[Effort and fee equilibrium]
\label{prop:feeeq}
Suppose
\begin{equation}
Z-\ell\alpha
\left[\gamma_\Delta\mathcal X(\Delta)
+\nu\delta_\Delta\mathcal R(\Delta)\right]>0,
\label{eq:feestability}
\end{equation}
Condition~\eqref{eq:concavity} holds, and the solution below satisfies partial coverage and Assumption~\ref{ass:trust}. A unique symmetric interior subgame-perfect equilibrium within the maintained regime is
\begin{align}
f^*&=\frac{\ell v+\lambda/2}
{Z-\ell\alpha[\gamma_\Delta\mathcal X(\Delta)+\nu\delta_\Delta\mathcal R(\Delta)]},
\label{eq:fstar}\\
x^*&=\mathcal X(\Delta)f^*,
\label{eq:xstar}\\
r^*&=\mathcal R(\Delta)f^*.
\label{eq:rstar}
\end{align}
Fund $B$ chooses $-r^*$.
\end{proposition}

Portfolio effort and fees are strategic complements in the symmetric partial-coverage equilibrium. Greater investor willingness to pay raises equilibrium fees, while a larger equilibrium fee increases the value of attracting an additional investor during the positioning stage.

\begin{corollary}[Disagreement, fees, and participation]
\label{cor:fees}
Under the conditions of Proposition~\ref{prop:feeeq}, disagreement lowers equilibrium fees whenever $c>0$ or $\nu<1$. Common effort, both systematic scores, and local participation also fall. Score-specialisation ratio
\begin{equation}
\frac{|d^*|}{g^*}
=\nu\frac{\phi_d}{\phi_g}\frac{\Delta}{2-\Delta}
\label{eq:specialisationratio}
\end{equation}
rises strictly with disagreement.
\end{corollary}

Preference-consistent alignment generates fee compression. Marginal fee return to common effort exceeds marginal fee return to provider-specific effort because contestable investors value only the common direction. Disagreement still makes portfolios more specialised in relative terms, but market segmentation does not raise fees without an additional investor-attention or salience channel.

\subsection{Cost incidence and local validity}
\label{sec:costincidence}

Baseline cost $\kappa(x_j^2+r_j^2)/2$ represents fund-level organisational resources. Financial-performance losses borne by each investor require a different incidence. Define per-account financial-performance loss as
\begin{equation}
\mathcal P_j=\frac{\beta}{2}(x_j^2+r_j^2),
\qquad \beta\geq0,
\label{eq:perunitloss}
\end{equation}
Loss $\mathcal P_j$ reduces investor utility and therefore demand for every invested account. Partial-coverage demand becomes
\begin{align}
D_j^{\beta}
={}&\ell\left[v-f_j+\alpha\gamma_\Delta x_j
+\alpha\nu_j\delta_\Delta r_j-\mathcal P_j\right]
+\frac{\lambda}{2} \\
&+c\left[\alpha\gamma_\Delta(x_j-x_k)
+\alpha\nu_C\delta_\Delta(r_j-r_k)
-(\mathcal P_j-\mathcal P_k)-(f_j-f_k)\right].
\label{eq:demandperformance}
\end{align}

\begin{proposition}[Investor-borne financial-performance loss]
\label{prop:performancecost}
Suppose every investor type has partial coverage and the fund earns fixed margin $m$. Each fund's unique interior effort choices under demand~\eqref{eq:demandperformance} are
\begin{equation}
x_j^{P,\beta}
=\frac{m\alpha b\gamma_\Delta}{\kappa+m\beta b},
\qquad
r_j^{P,\beta}
=\frac{m\alpha\zeta_j\delta_\Delta}{\kappa+m\beta b}.
\label{eq:performanceeffort}
\end{equation}
Investor-borne losses attenuate both effort directions, preserve $\operatorname{sign}(r_j)=\operatorname{sign}(\zeta_j)$, and leave relative score specialisation
\begin{equation}
\frac{|d_j^{P,\beta}|}{g_j^{P,\beta}}
=\frac{|\zeta_j|}{b}\frac{\Delta}{2-\Delta}
\label{eq:performanceratio}
\end{equation}
strictly increasing in disagreement whenever $\zeta_j\neq0$. A fund's favoured systematic score is non-increasing in $\Delta$ because $|\zeta_j|\leq b$.
\end{proposition}

Manager-level and investor-borne costs represent distinct economic objects. Manager-level cost limits organisational implementation, whereas investor-borne loss reduces demand for financially costly portfolio deviations. Joint incidence preserves the sign of provider targeting and the increase in relative specialisation.

Assumption~\ref{ass:trust} makes the core rating technology exact on the feasible affine cell. Fixed-margin candidates require
\begin{equation}
(x_P)^2+(r_P)^2<R^2,
\label{eq:trustfixed}
\end{equation}
while endogenous-fee candidates require
\begin{equation}
(x^*)^2+(r^*)^2<R^2.
\label{eq:trustfee}
\end{equation}
Appendix~\ref{sec:nonlinearrobustness} shows that strong concavity bounds the best-response error created by replacing the exact affine rating map with a smooth nonlinear map.

Proposition~\ref{prop:feeeq} characterises fee competition in the economically central partial-coverage regime. Full-coverage and boundary fee games are piecewise because a unilateral fee deviation can change market coverage. Subsequent results do not extrapolate the partial-coverage fee solution across coverage boundaries.

\section{Rating systems and investor reliance}
\label{sec:ratingsystems}

A two-methodology baseline makes the common-versus-differential portfolio technology transparent. A larger provider set preserves the production-cost decomposition, while endogenous attention can amplify investor sorting across provider methodologies.

\subsection{Portfolio technology with multiple providers}
\label{sec:manyproviders}

Let $s\in\R^J$ collect systematic scores from $J$ providers and let $G=BH^{-1}B'$ be the methodology Gram matrix. Full row rank requires $J\leq K-1$. Lemma~\ref{lem:implementation} gives cost $\kappa s'G^{-1}s/2$. Suppose methodologies have common pairwise similarity $\rho\in(-1/(J-1),1)$:
\begin{equation}
G=(1-\rho)I_J+\rho\ind\ind'.
\label{eq:equicorrelation}
\end{equation}
Decompose scores into a common component and an orthogonal provider-specific component:
\begin{equation}
s=g\ind+s^\perp,
\qquad
\ind's^\perp=0.
\label{eq:multidecomp}
\end{equation}

\begin{proposition}[Common and provider-specific production costs]
\label{prop:many}
Under \eqref{eq:equicorrelation}, minimum portfolio cost is
\begin{equation}
C_J(g,s^\perp)=\frac{\kappa}{2}
\left[
\frac{Jg^2}{1+(J-1)\rho}
+\frac{\lVert s^\perp\rVert^2}{1-\rho}
\right].
\label{eq:manycost}
\end{equation}
Greater methodology similarity lowers the cost of the common score direction and raises the cost of every provider-specific score direction.
\end{proposition}

General methodology matrices admit eigendecomposition $G=V\Lambda V'$. Let $a_s$ collect the exogenous marginal demand returns to the $J$ provider scores. A fund chooses
\begin{equation}
s^*=\frac{m}{\kappa}Ga_s
=\frac{m}{\kappa}V\Lambda V'a_s.
\label{eq:manysolution}
\end{equation}
Equation~\eqref{eq:manysolution} generalises the portfolio-production technology rather than solving a complete strategic equilibrium with $J$ providers. Investor sorting and fee competition would determine marginal demand-return vector $a_s$ endogenously.

\subsection{Endogenous investor reliance}
\label{sec:attention}

Baseline clienteles have fixed mean alignment. Heterogeneous investor priorities allow methodology disagreement to change whether investors process a provider-specific methodology or rely on a balanced composite. Investor $i$'s preferred methodology vector is $\gamma_\Delta u+\nu_i\delta_\Delta v$, while the balanced composite uses $\gamma_\Delta u$. Squared mismatch under metric $H^{-1}$ equals $\nu_i^2\delta_\Delta^2$. Ignoring provider-specific information creates the reduced-form mismatch loss
\begin{equation}
L_i(\Delta)=\frac{\omega}{2}\nu_i^2\delta_\Delta^2
=\frac{\omega}{4}\nu_i^2\Delta,
\label{eq:attentionloss}
\end{equation}
where $\omega>0$ scales the value of matching investor priorities.

Let $F_k(\cdot\mid\nu_i)$ denote the conditional distribution of attention cost $k_i$. Assume a continuously differentiable conditional distribution with density $f_k(\cdot\mid\nu_i)$ and sufficient domination to interchange differentiation and integration. Investor $i$ follows the provider whose methodology is closer to the investor's preferred vector when $L_i(\Delta)\geq k_i$. Average provider-specific alignment among attentive investors in each symmetric local clientele becomes
\begin{equation}
\nu(\Delta)=\E\left[
|\nu_i|F_k\left(\frac{\omega}{4}\nu_i^2\Delta\middle|\nu_i\right)
\middle|\text{local clientele}
\right],
\label{eq:nudelta}
\end{equation}
Composite users contribute zero to $\nu(\Delta)$, whereas attentive users contribute their absolute provider alignment. Differentiating the attentive-alignment index gives
\begin{equation}
\nu'(\Delta)=\frac{\omega}{4}\E\left[
|\nu_i|^3 f_k\left(\frac{\omega}{4}\nu_i^2\Delta\middle|\nu_i\right)
\middle|\text{local clientele}
\right]\geq0.
\label{eq:nudeltaprime}
\end{equation}
Fixed-margin partial equilibrium becomes
\begin{equation}
x_P(\Delta)=\frac{m\alpha b\gamma_\Delta}{\kappa},
\qquad
r_P(\Delta)=\frac{m\alpha\ell\nu(\Delta)\delta_\Delta}{\kappa}.
\label{eq:attentioneffort}
\end{equation}

\begin{proposition}[Endogenous provider reliance]
\label{prop:attention}
Under the differentiability conditions above, a favoured systematic score rises with disagreement if and only if
\begin{equation}
\ell\left[\nu(\Delta)+\Delta\nu'(\Delta)\right]>b.
\label{eq:attentionscore}
\end{equation}
Total participation rises if and only if
\begin{equation}
\ell\left[\nu(\Delta)^2
+2\Delta\nu(\Delta)\nu'(\Delta)\right]>b.
\label{eq:attentionparticipation}
\end{equation}
Common portfolio effort continues to fall.
\end{proposition}

When $\nu'(\Delta)=0$, neither threshold can hold if contestable demand is positive. A sufficiently rapid increase in provider reliance can reverse the baseline decline in a favoured score or participation because disagreement changes both score productivity and investor information processing. Equation~\eqref{eq:attentionloss} derives the reliance incentive from the same portfolio geometry used in the production model.

\section{Welfare and methodology convergence}
\label{sec:welfare}

Reported scores, flows, and fees do not provide a welfare ranking. Fee payments transfer surplus between investors and funds. Portfolio deviations consume financial resources. Holdings-based outcome $y$ enters separately from either reported rating.

\subsection{Welfare at the endogenous-fee equilibrium}
\label{sec:welfareeq}

Consider the symmetric partial-coverage equilibrium in Proposition~\ref{prop:feeeq}. Attached distributions are reflections of one another: $F_B(\nu)=F_A(-\nu)$, with means $\nu_A=\nu$ and $\nu_B=-\nu$. Contestable alignment has mean zero. Define
\begin{equation}
\sigma_{\nu,L}^2=\Var_{F_A}(\nu_i)=\Var_{F_B}(\nu_i),
\qquad
\mu_{\nu,C}^{(2)}=\E_{F_C}[\nu_i^2].
\label{eq:alignmentmoments}
\end{equation}
Support-wide interiority in Assumption~\ref{ass:contest} makes first moments sufficient for demand, but welfare also depends on alignment variance in attached segments and the second moment of contestable alignment.

Gross local value before fees and non-ESG mismatch, evaluated at mean attached alignment, is
\begin{equation}
q_0=v+\alpha\gamma_\Delta x+\alpha\nu\delta_\Delta r.
\label{eq:q0}
\end{equation}
Total participation is
\begin{equation}
Q(x,r,f;\Delta)=\lambda+
\frac{1-\lambda}{t}(q_0-f).
\label{eq:Q}
\end{equation}
Equal demands and opposite tilts make average holdings-based outcome $\theta_u x$. Welfare, up to constants independent of choices, is
\begin{align}
\mathcal W(x,r,f;\Delta)
={}&\frac{1-\lambda}{2t}
\left[q_0^2-f^2
+\alpha^2\delta_\Delta^2r^2\sigma_{\nu,L}^2\right]
+\lambda\alpha\gamma_\Delta x \notag\\
&+\frac{\lambda\alpha^2\delta_\Delta^2r^2}{h}
\mu_{\nu,C}^{(2)}
-\kappa(x^2+r^2)
+\eta\theta_u xQ(x,r,f;\Delta),
\label{eq:welfare}
\end{align}
where $\eta\geq0$ measures the social value assigned to holdings-based exposure. Alignment variance $\sigma_{\nu,L}^2$ captures matching within attached segments. Contestable second moment $\mu_{\nu,C}^{(2)}$ captures gains from sorting contestable investors between oppositely tilted funds. Both matching terms disappear when investor alignment is degenerate.

\begin{proposition}[Welfare decomposition with heterogeneous priorities]
\label{prop:welfare}
Let $\mathcal W^*(\Delta)=\mathcal W(x^*(\Delta),r^*(\Delta),f^*(\Delta);\Delta)$. Then
\begin{equation}
\frac{d\mathcal W^*}{d\Delta}
=\mathcal W_x\frac{dx^*}{d\Delta}
+\mathcal W_r\frac{dr^*}{d\Delta}
+\mathcal W_f\frac{df^*}{d\Delta}
+\mathcal W_\Delta,
\label{eq:Wdecomp}
\end{equation}
where
\begin{align}
\mathcal W_x
&=\frac{1-\lambda}{t}\alpha\gamma_\Delta q_0
+\lambda\alpha\gamma_\Delta-2\kappa x
+\eta\theta_u\left[Q+
\frac{1-\lambda}{t}\alpha\gamma_\Delta x\right],
\label{eq:Wx}\\
\mathcal W_r
&=\frac{1-\lambda}{t}\alpha\nu\delta_\Delta q_0
+\frac{1-\lambda}{t}\alpha^2\delta_\Delta^2r\sigma_{\nu,L}^2
+\frac{2\lambda\alpha^2\delta_\Delta^2r}{h}\mu_{\nu,C}^{(2)}
-2\kappa r \notag\\
&\quad+\eta\theta_u\frac{1-\lambda}{t}
\alpha\nu\delta_\Delta x,
\label{eq:Wr}\\
\mathcal W_f
&=-\frac{1-\lambda}{t}(f+\eta\theta_u x).
\label{eq:Wf}
\end{align}
Holding choices fixed, define
\begin{equation}
q_\Delta=\alpha\gamma_\Delta' x
+\alpha\nu\delta_\Delta' r.
\label{eq:qDelta}
\end{equation}
Holding portfolio efforts and fees fixed, the direct derivative with respect to methodology disagreement is
\begin{align}
\mathcal W_\Delta
={}&\frac{1-\lambda}{t}(q_0+\eta\theta_u x)q_\Delta
+\lambda\alpha\gamma_\Delta' x \notag\\
&+\alpha^2r^2(\delta_\Delta^2)'
\left[\frac{1-\lambda}{2t}\sigma_{\nu,L}^2
+\frac{\lambda}{h}\mu_{\nu,C}^{(2)}\right].
\label{eq:WDelta}
\end{align}
\end{proposition}

Effort coordinates keep outcome loadings $\theta_u$ and $\theta_v$ fixed as methodologies rotate. Symmetric demands and opposite tilts remove provider-specific outcome component $\theta_v r$ from aggregate exposure, but provider-specific effort still creates matching surplus. Unequal demands restore a provider-specific contribution to aggregate holdings-based exposure:
\begin{equation}
Y=D_A(\theta_u x_A+\theta_v r_A)
+D_B(\theta_u x_B+\theta_v r_B).
\label{eq:asymoutcome}
\end{equation}

Closed-form welfare comparison follows from the fee equilibrium. Define
\begin{align}
\mathcal N&=\ell v+\lambda/2,
& \mathcal H(\Delta)&=\frac{b\alpha^2}{\kappa}
\left[\phi_g(2-\Delta)+\nu^2\phi_d\Delta\right],
\label{eq:welfareaux1}\\
\mathcal D(\Delta)&=Z-\ell\mathcal H(\Delta),
& f^*(\Delta)&=\frac{\mathcal N}{\mathcal D(\Delta)},
\label{eq:welfareaux2}\\
K_x&=\frac{2b\alpha \phi_g}{\kappa},
& K_r&=\frac{2b\alpha\nu \phi_d}{\kappa},
\label{eq:welfareaux3}\\
\mathcal J_\nu&=\ell\sigma_{\nu,L}^2
+\frac{\lambda}{h}\mu_{\nu,C}^{(2)},
& \rho_f(\Delta)&=\frac{f^{*\prime}}{f^*}
=\frac{\ell\mathcal H'(\Delta)}{\mathcal D(\Delta)}<0.
\label{eq:welfareaux4}
\end{align}
Equilibrium choices satisfy $x^*=K_x\gamma_\Delta f^*$, $r^*=K_r\delta_\Delta f^*$, and $q_0^*=v+\mathcal H(\Delta)f^*$. Private surplus, excluding the holdings-based term, is
\begin{align}
V^*(\Delta)
={}&\ell\left[(q_0^*)^2-(f^*)^2\right]
+\lambda\alpha K_x\gamma_\Delta^2 f^* \notag\\
&+\mathcal J_\nu\alpha^2K_r^2\delta_\Delta^4(f^*)^2
-\kappa(f^*)^2
\left(K_x^2\gamma_\Delta^2+K_r^2\delta_\Delta^2\right).
\label{eq:privatewelfare}
\end{align}
Holdings-based exposure equals
\begin{equation}
Y^*(\Delta)=2b\theta_uK_x\gamma_\Delta(f^*)^2.
\label{eq:Yclosed}
\end{equation}

\begin{proposition}[Closed-form outcome-value threshold]
\label{prop:welfarethreshold}
Under Proposition~\ref{prop:feeeq}, Assumption~\ref{ass:contest}, and $\theta_u>0$, define $q_0^*=v+\mathcal H(\Delta)f^*$ and
\begin{align}
V^{*\prime}(\Delta)
={}&2\ell f^*\left[q_0^*\{\mathcal H'(\Delta)+\mathcal H(\Delta)\rho_f(\Delta)\}-f^*\rho_f(\Delta)\right]
+\lambda\alpha K_x f^*
\left[-\frac12+\gamma_\Delta^2\rho_f(\Delta)\right] \notag\\
&+\mathcal J_\nu\alpha^2K_r^2(f^*)^2
\left[\delta_\Delta^2+2\delta_\Delta^4\rho_f(\Delta)\right] \notag\\
&-\kappa(f^*)^2\left[
K_x^2\left\{-\frac12+2\gamma_\Delta^2\rho_f(\Delta)\right\}
+K_r^2\left\{\frac12+2\delta_\Delta^2\rho_f(\Delta)\right\}
\right].
\label{eq:Vprimeclosed}
\end{align}
Exposure derivative is
\begin{equation}
Y^{*\prime}(\Delta)
=Y^*(\Delta)
\left[-\frac{1}{2(2-\Delta)}+2\rho_f(\Delta)\right]<0.
\label{eq:Yprimeclosed}
\end{equation}
Set
\begin{equation}
\bar\eta(\Delta)=
\max\left\{0,
\frac{V^{*\prime}(\Delta)}{-Y^{*\prime}(\Delta)}
\right\}.
\label{eq:etathreshold}
\end{equation}
A marginal reduction in methodology disagreement raises welfare whenever $V^{*\prime}(\Delta)\leq0$ or $\eta>\bar\eta(\Delta)$. If $V^{*\prime}(\Delta)>0$, condition $\eta>\bar\eta(\Delta)$ is also necessary. Numerator $V^{*\prime}(\Delta)$ and denominator $-Y^{*\prime}(\Delta)$ are closed-form functions of the model primitives.
\end{proposition}

Provider-specific effort can raise private surplus by matching heterogeneous priorities even while common holdings exposure falls. Equation~\eqref{eq:Vprimeclosed} collects attached and contestable matching gains in coefficient $J$. Greater dispersion in investor priorities can therefore raise the social value of common exposure required for methodology convergence to improve welfare.

\subsection{Methodology convergence}
\label{sec:convergence}

Several policies affect distinct model primitives. Harmonised corporate disclosure can reduce measurement variances without changing methodology gradients. Common taxonomies can change both methodology directions and measurement precision. Convergence in portfolio-rating methodologies lowers $\Delta$. Standardised rating scales improve comparability without necessarily changing any portfolio direction.

Methodology convergence raises common portfolio effort, both reported systematic scores, participation, and equilibrium fees in the preference-consistent symmetric benchmark. Under the fixed-bisector path, convergence also raises average holdings-based outcome when $\theta_u>0$. Proposition~\ref{prop:welfarethreshold} identifies when the increase in common holdings-based exposure dominates changes in private surplus.

A rotating methodology bisector changes the relationship between disagreement and holdings-based exposure. Let common direction become $u(\Delta)$ and define $\theta_u(\Delta)=a'H^{-1}u(\Delta)$. Aggregate symmetric exposure then satisfies
\begin{equation}
\frac{dY^*}{d\Delta}
=\theta_u' x^*Q^*+\theta_u\left(x^{*\prime}Q^*+x^*Q^{*\prime}\right).
\label{eq:rotatingoutcome}
\end{equation}
Under a rotating bisector, disagreement still lowers aggregate holdings-based exposure whenever rotation toward outcome vector $a$ is not strong enough to offset falling common effort and participation. Policy conclusions therefore apply to convergence that changes relative methodology weights while preserving common direction $u$. Disclosure reform, taxonomy changes, or scope expansion may rotate $u$ and require equation~\eqref{eq:rotatingoutcome} instead.

\section{Economic implications}
\label{sec:implications}

Equilibrium and welfare propositions generate a focused set of comparative statics. Numerical illustrations verify the maintained regime and show how stronger disagreement changes effort, fees, participation, and welfare.

\subsection{Numerical comparative statics}
\label{sec:numerical}

Baseline illustrations use $t=\kappa=1$, $m=0.8$, $\alpha=0.5$, $\lambda=0.3$, $h=0.6$, $\nu=0.75$, $v=0.4$, $\theta_u=0.8$, and trust radius $R=0.4$, with $\Delta\in[0.05,0.95]$. Fund $A$'s attached alignments are uniform on $[0.55,0.95]$, while Fund $B$'s attached alignments are uniform on $[-0.95,-0.55]$; both distributions have variance $\sigma_{\nu,L}^2=1/75$. Contestable alignments are uniform on $[-0.5,0.5]$, giving $\mu_{\nu,C}^{(2)}=1/12$. Every plotted equilibrium satisfies support-wide partial coverage, support-wide contestable interiority, strict trust-region validity, a positive fee denominator, and stage-one concavity.

\begin{figure}[H]
\centering
\includegraphics[width=0.86\textwidth]{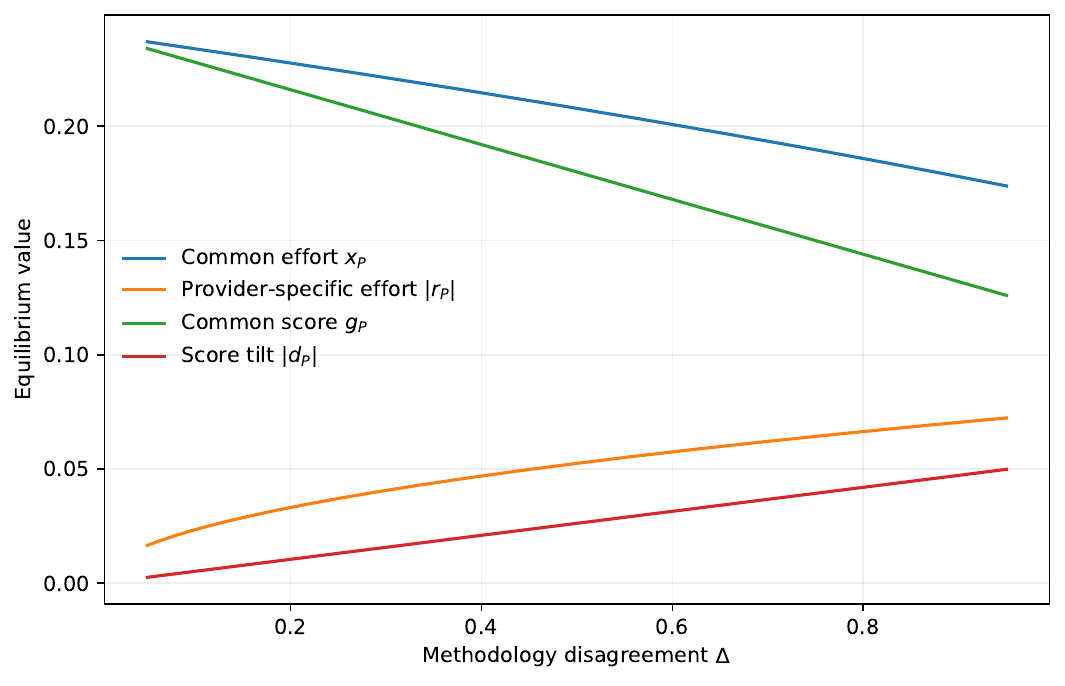}
\caption{Systematic disagreement, portfolio effort, and score components. Common effort $x$ and common score $g$ decline, while provider-specific effort $|r|$ and absolute score tilt $|d|$ rise.}
\label{fig:effort}
\end{figure}

Figure~\ref{fig:effort} plots fixed-margin effort and score components. Common portfolio effort and common score component fall. Provider-specific effort and absolute score tilt rise. Both reported ratings nevertheless fall because provider-specific valuation remains bounded by total ESG valuation.

\begin{figure}[H]
\centering
\includegraphics[width=0.86\textwidth]{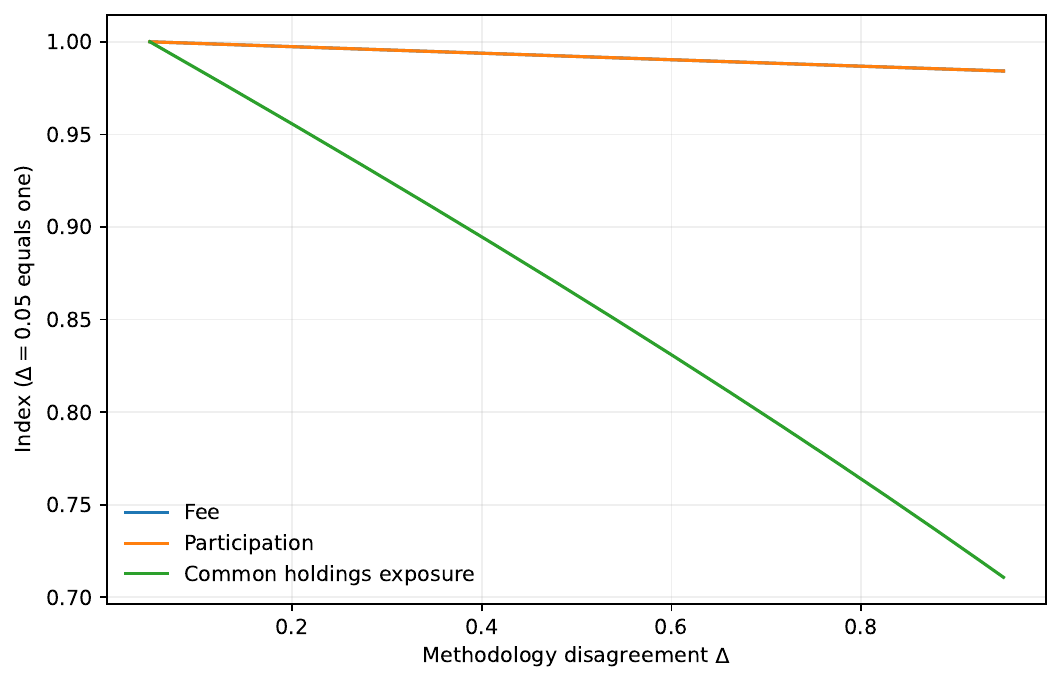}
\caption{Endogenous-fee equilibrium, normalised to one at $\Delta=0.05$. Fees, total participation, and common holdings-based exposure fall with systematic disagreement.}
\label{fig:fees}
\end{figure}

Figure~\ref{fig:fees} evaluates the endogenous-fee equilibrium. Normalised fees, participation, and common holdings-based exposure decline with disagreement. Relative score specialisation nevertheless rises because disagreement increases the score productivity of provider-specific effort relative to common effort, as equation~\eqref{eq:specialisationratio} shows.

\begin{figure}[H]
\centering
\includegraphics[width=0.86\textwidth]{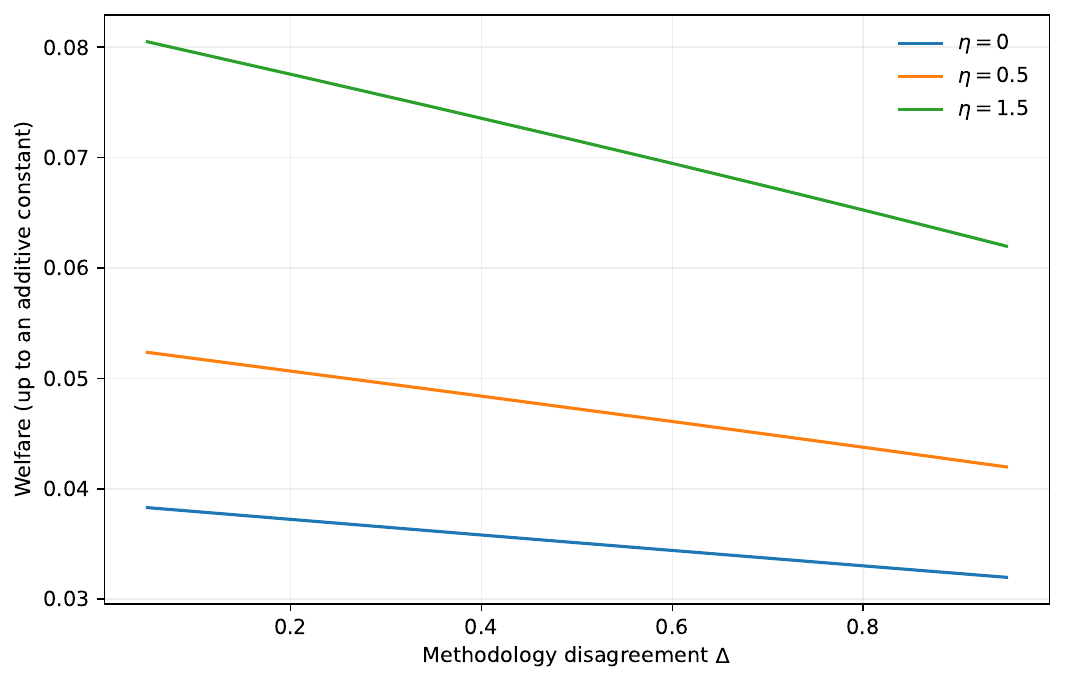}
\caption{Welfare under systematic disagreement for $\eta\in\{0,0.5,1.5\}$. Private surplus includes matching gains from heterogeneous investor priorities.}
\label{fig:welfare}
\end{figure}

Figure~\ref{fig:welfare} reports welfare for three values of outcome weight $\eta$. Private welfare includes matching gains from attached and contestable preference dispersion. Stronger outcome valuation increases the welfare cost of disagreement because common holdings-based exposure falls. Curves remain illustrative rather than calibrated.

\subsection{Empirical implications}
\label{sec:empirical}

Cross-sectional variation in investor priorities predicts the sign of provider targeting. Funds serving similarly aligned investors should tilt in the same methodology direction. Funds serving oppositely aligned investors should report widening score spreads in opposite directions as disagreement rises. Distribution channels, adviser affiliations, and retirement-plan menus can identify persistent local clienteles, while holdings reveal signed portfolio adjustments.

Portfolio constraints determine the magnitude of portfolio adjustment. Stronger benchmark discipline, narrower investment universes, and costly provider-specific attributes reduce both common and differential effort. \cite{gantchev2024sustainability} document portfolio changes undertaken to improve sustainability ratings and the associated performance cost . Passive mandates and tightly benchmarked active funds should therefore adjust less than unconstrained funds.

Methodology changes provide sharper identifying variation. Mechanical issuer-rating upgrades increase security selection by ESG funds even without changes in underlying activity \citep{choi2026mechanical}. Model predictions distinguish common holdings that improve both methodologies from signed reallocation toward one provider. Greater systematic divergence should reduce common-direction holdings and increase the absolute projection of portfolio changes onto the differential direction.

Preference-consistent alignment yields a distinctive flow prediction. Stronger disagreement widens score dispersion but lowers both provider scores and aggregate participation unless investor attention changes endogenously. Flow growth following divergence would therefore indicate sorting or salience beyond fixed convex-combination preferences. Proposition~\ref{prop:attention} identifies the required attention margin.

Trading activity supplies a separate participation test. Using MSCI-IVA, LSEG-ESG, and Sustainalytics, \citet{quotb2026trading} associate a one-standard-deviation increase in disagreement with a 1.3\% decline in abnormal trading volume. \citet{quotb2026trading} report no significant disagreement--volume relation during 2009--2016 but a negative relation during 2017--2022, consistent with growing investor salience. Stronger responses among norm-constrained institutions accord with heterogeneous ESG priorities, while wider bid--ask spreads indicate trading frictions beyond the fund-choice mechanism modelled here.

Fee-response sign distinguishes competing economic channels. Baseline disagreement compresses fees because common willingness to pay falls faster than market segmentation rises. Higher fees following greater disagreement would indicate stronger provider-specific salience, endogenous investor sorting, or asymmetric distribution advantages absent from the benchmark. Empirical analysis should therefore examine fees jointly with score spreads, holdings, flows, trading activity, and liquidity.

Empirical implementation must distinguish four quantities. Provider scores identify common and differential score components. Portfolio holdings identify common and differential effort. Holdings-based environmental or social measures identify outcome loadings $\theta_u$ and $\theta_v$. Rating revisions unexplained by holdings identify measurement uncertainty. Score specialisation alone is insufficient evidence of greenwashing because genuine portfolio changes can improve one methodology while worsening another.

\begin{table}[t]
\centering
\caption{Main predictions}
\label{tab:predictions}
\small
\begin{tabularx}{\textwidth}{>{\raggedright\arraybackslash}p{0.28\textwidth}>{\raggedright\arraybackslash}p{0.29\textwidth}>{\raggedright\arraybackslash}X}
\toprule
Change & Equilibrium response & Observable implication \\
\midrule
More systematic disagreement & Lower common effort; greater absolute signed tilt & Lower common holdings factor; larger cross-provider score spread \\
Oppositely aligned clienteles & Opposite signs of provider-specific effort & Funds target different methodologies without an imposed provider assignment \\
More contestable investors & Weaker relative specialisation; common effort rises only when $h<t$ & Smaller score dispersion; common holdings response depends on relative differentiation parameters \\
Higher implementation cost & Lower effort in both portfolio directions & Weaker holdings response among implementation-constrained funds \\
Greater measurement uncertainty & Lower participation for positive expected utility; contestable elasticity falls, while local elasticity falls near the participation margin & More dispersed ratings and a weaker flow response where investors remain close to participation or switching margins \\
Stronger endogenous attention & Faster growth in provider-specific effort & Favoured scores or flows can rise only when sorting crosses Proposition~\ref{prop:attention}'s threshold \\
\bottomrule
\end{tabularx}
\end{table}

\section{Conclusion}
\label{sec:conclusion}

Conflicting ESG rating methodologies reshape fund portfolios before published ratings guide investor allocation. Explicit methodology geometry separates common portfolio effort from signed provider-specific effort. Funds choose the direction of specialisation according to their clienteles' ESG priorities. Opposite provider targeting therefore emerges from opposite investor alignment rather than from an imposed portfolio restriction.

Greater systematic disagreement reduces the score productivity of common effort and increases the productivity of provider-specific effort. Within the partial-coverage benchmark, equilibrium portfolios become more specialised, yet both systematic ratings, participation, and fees fall under preference-consistent investor weights. Endogenous attention can reverse the declines in scores, participation, and fees only when disagreement induces sufficiently rapid investor sorting.

Holdings-based ESG outcomes remain distinct from provider scores. Fixed portfolio directions make methodology convergence mathematically transparent and avoid treating score averages as truth. Preference dispersion also gives provider-specific portfolios a matching value that demand means alone cannot reveal. Manager-level implementation costs remain distinct from investor-borne financial-performance losses, while nonlinear rating maps generate nearby equilibria under a contraction condition. Policy conclusions must therefore distinguish disclosure quality, methodology convergence, rating precision, and rating-scale standardisation because each intervention changes a different model primitive.

Competing methodologies matter not only because ratings transmit information. Methodology differences also alter the production technology of financial products, the direction of portfolio differentiation, and the price at which funds serve heterogeneous investors.

\bibliography{funds2.bib}
\bibliographystyle{apalike}

\appendix

\section{Proofs}
\label{app:proofs}

\subsection{Portfolio technology}

\paragraph{Proof of Lemma~\ref{lem:implementation}.}
Form the Lagrangian
\begin{equation}
\mathcal J(z,\vartheta)=\frac{\kappa}{2}z'Hz+\vartheta'(s-Bz).
\end{equation}
First-order condition $\kappa Hz-B'\vartheta=0$ gives $z=H^{-1}B'\vartheta/\kappa$. Constraint $Bz=s$ implies $\vartheta=\kappa(BH^{-1}B')^{-1}s$. Substitution gives \eqref{eq:minportfolio}. Evaluating \eqref{eq:financialloss} at minimum-cost portfolio $z^*$ yields \eqref{eq:generalcost}; substitution of $z^*$ into $y=a'z$ yields \eqref{eq:outcomemap}. \qed

Equations~\eqref{eq:scoresxr}--\eqref{eq:costoutcomexr} follow directly from \eqref{eq:uvorthogonal}--\eqref{eq:effortportfolio}. In particular,
\begin{equation}
b_1'z=\gamma_\Delta x+\delta_\Delta r,
\qquad
b_2'z=\gamma_\Delta x-\delta_\Delta r,
\end{equation}
and
\begin{equation}
z'Hz=(xu+rv)'H^{-1}(xu+rv)=x^2+r^2.
\end{equation}
Substituting $x=g/\gamma_\Delta$ and $r=d/\delta_\Delta$ yields \eqref{eq:scorecost}.

\subsection{Positioning and coverage}

\paragraph{Proof of Proposition~\ref{prop:direction}.}
Under partial coverage, Fund $j$ demand is
\begin{equation}
D_j=\ell\left(v+\alpha\gamma_\Delta x_j
+\alpha\nu_j\delta_\Delta r_j\right)
+\frac{\lambda}{2}
+c\left[\alpha\gamma_\Delta(x_j-x_k)
+\alpha\nu_C\delta_\Delta(r_j-r_k)\right].
\end{equation}
Profit is strictly concave in $(x_j,r_j)$. First-order conditions are
\begin{equation}
m\alpha b\gamma_\Delta-\kappa x_j=0,
\qquad
m\alpha(\ell\nu_j+c\nu_C)\delta_\Delta-\kappa r_j=0.
\end{equation}
Solving gives \eqref{eq:generalpartial}. Positive proportionality between $r_j$ and $\nu_j$ proves \eqref{eq:signdirection}. \qed

\paragraph{Proof of Proposition~\ref{prop:regimes}.}
Partial candidate follows Proposition~\ref{prop:direction}. Full coverage fixes local demand. Remaining marginal return is $mc\alpha\gamma_\Delta$, so strict concavity gives \eqref{eq:fullcandidate}.

Suppose $q_F<t\leq q_P$. Coverage boundary satisfies $a_xx+a_rr=R_t$. Variable profit along the coverage boundary is
\begin{equation}
mc\alpha\gamma_\Delta x-\frac{\kappa}{2}(x^2+r^2).
\end{equation}
Unconstrained full-coverage point is $(x_F,0)$. Orthogonal projection of $(x_F,0)$ onto $a_xx+a_rr=R_t$ gives \eqref{eq:boundaryx}--\eqref{eq:boundaryr}.

Positive local participation makes the participation rule equal to the pointwise minimum of a linear function and one, which is concave. Adding linear contestable demand and subtracting strictly convex portfolio cost makes profit strictly concave over the convex trust region. Regime-valid candidate choices therefore identify the unique global maximiser within the local model. \qed

\paragraph{Proof of Proposition~\ref{prop:reallocation}.}
Differentiate \eqref{eq:gammadelta}:
\begin{equation}
\gamma_\Delta'=-\frac{1}{4\gamma_\Delta},
\qquad
\delta_\Delta'=\frac{1}{4\delta_\Delta}.
\end{equation}
Substitution into \eqref{eq:partialcandidate} gives \eqref{eq:xrderivatives}. Since $\gamma_\Delta^2=(2-\Delta)/2$ and $\delta_\Delta^2=\Delta/2$, equation~\eqref{eq:gdpartial} gives \eqref{eq:gdderivatives}. Adding and subtracting the derivatives of $g_P$ and $d_P$ yields \eqref{eq:favouredfall}--\eqref{eq:unfavouredfall}. Local gross utility is
\begin{equation}
q_P=v+\frac{m\alpha^2}{2\kappa}
\left[b(2-\Delta)+\nu^2\ell\Delta\right].
\end{equation}
Differentiation gives \eqref{eq:qfall}. \qed

\subsection{Fees, cost incidence, and welfare}

\paragraph{Proof of Lemma~\ref{lem:fee}.}
Fund $A$'s fee first-order condition is $A_A-2bf_A+cf_B=0$. Fund $B$ has the symmetric condition. Solving the linear system gives \eqref{eq:feeA}--\eqref{eq:feeB}. Symmetry gives \eqref{eq:symmetricfee}; fee optimality also gives $D=bf$. \qed

\paragraph{Proof of Proposition~\ref{prop:feeeq}.}
At fee equilibrium, operating profit is $bf_A^2$. Fee sensitivities are
\begin{equation}
\frac{\partial f_A}{\partial x_A}=\alpha\gamma_\Delta \phi_g,
\qquad
\frac{\partial f_A}{\partial r_A}=\alpha\nu_A\delta_\Delta \phi_d.
\end{equation}
Stage-one first-order conditions at a symmetric allocation are
\begin{equation}
2bf\alpha\gamma_\Delta \phi_g-\kappa x=0,
\qquad
2bf\alpha\nu\delta_\Delta \phi_d-\kappa r=0.
\end{equation}
Stage-one first-order conditions give $x=\mathcal Xf$ and $r=\mathcal Rf$. Substitution into \eqref{eq:symmetricfee} yields \eqref{eq:fstar}.

Write $q_f=(\alpha\gamma_\Delta \phi_g,\alpha\nu\delta_\Delta \phi_d)'$. Stage-one Hessian is $2bq_fq_f'-\kappa I_2$. Negative definiteness is equivalent to $2bq_f'q_f<\kappa$, which is \eqref{eq:concavity}. Strict concavity and positive denominator deliver the unique symmetric interior equilibrium within the maintained regime. \qed

\paragraph{Proof of Corollary~\ref{cor:fees}.}
Define
\begin{equation}
A(\Delta)=\frac{2b\alpha^2}{\kappa}
\left(\phi_g\gamma_\Delta^2+\nu^2\phi_d\delta_\Delta^2\right).
\end{equation}
Equation~\eqref{eq:fstar} is $f^*=(\ell v+\lambda/2)/(Z-\ell A)$. Derivative $A'$ has the sign of $\nu^2\phi_d-\phi_g$. Since
\begin{equation}
\phi_g-\phi_d=\frac{c(2b-c)}{M}>0
\end{equation}
whenever $c>0$, and $\nu^2\leq1$, $A'<0$ whenever $c>0$ or $\nu<1$. Hence $f^{*\prime}<0$.

Common effort $x^*=\mathcal Xf^*$ falls because both $\gamma_\Delta$ and $f^*$ fall. Score components satisfy
\begin{equation}
g^*=\frac{b\alpha \phi_g}{\kappa}(2-\Delta)f^*,
\qquad
d^*=\frac{b\alpha\nu \phi_d}{\kappa}\Delta f^*.
\end{equation}
Both $g^*+d^*$ and $g^*-d^*$ fall because $f^{*\prime}<0$, $\nu \phi_d-\phi_g<0$, and $-\nu \phi_d-\phi_g<0$. Net local utility can be written as a strictly increasing function of $A$:
\begin{equation}
q_0-f=v+(A-1)\frac{\ell v+\lambda/2}{Z-\ell A},
\end{equation}
whose derivative with respect to $A$ is $(\ell v+\lambda/2)(Z-\ell)/(Z-\ell A)^2>0$. Participation therefore falls as $A$ falls. Ratio \eqref{eq:specialisationratio} follows by dividing $d^*=\delta_\Delta r^*$ by $g^*=\gamma_\Delta x^*$. \qed

\paragraph{Proof of Proposition~\ref{prop:performancecost}.}
Differentiate profit $mD_j^\beta-\kappa(x_j^2+r_j^2)/2$. Own demand derivatives are
\begin{equation}
\frac{\partial D_j^\beta}{\partial x_j}
=\alpha b\gamma_\Delta-\beta b x_j,
\qquad
\frac{\partial D_j^\beta}{\partial r_j}
=\alpha\zeta_j\delta_\Delta-\beta b r_j.
\end{equation}
Strict concavity follows from $\kappa+m\beta b>0$. Solving the first-order conditions gives \eqref{eq:performanceeffort}. Score ratio \eqref{eq:performanceratio} follows from $g=\gamma_\Delta x$ and $d=\delta_\Delta r$. Favoured score derivative is proportional to $-b+|\zeta_j|\leq0$ because $|\nu_j|,|\nu_C|\leq1$. \qed

\subsection{Rating systems and welfare}

\paragraph{Proof of Proposition~\ref{prop:many}.}
Matrix~\eqref{eq:equicorrelation} has eigenvalue $1+(J-1)\rho$ in common direction $\ind$ and eigenvalue $1-\rho$ in every orthogonal direction. Since $s=g\ind+s^\perp$ and $\ind's^\perp=0$,
\begin{equation}
s'G^{-1}s
=\frac{Jg^2}{1+(J-1)\rho}
+\frac{\lVert s^\perp\rVert^2}{1-\rho}.
\end{equation}
Lemma~\ref{lem:implementation} yields \eqref{eq:manycost}. \qed

\paragraph{Proof of Proposition~\ref{prop:attention}.}
Conditional attention probability is
\begin{equation}
F_k\left(\frac{\omega}{4}\nu_i^2\Delta\middle|\nu_i\right).
\end{equation}
Dominated differentiation gives equation~\eqref{eq:nudeltaprime}. Reported common component is
\begin{equation}
g_P=\frac{m\alpha b(2-\Delta)}{2\kappa},
\end{equation}
and absolute tilt is
\begin{equation}
d_P=\frac{m\alpha\ell\Delta\nu(\Delta)}{2\kappa}.
\end{equation}
Differentiation gives favoured-score condition~\eqref{eq:attentionscore}. Local utility contribution from tilt is
\begin{equation}
\alpha\nu(\Delta)d_P
=\frac{m\alpha^2\ell}{2\kappa}\Delta\nu(\Delta)^2.
\end{equation}
Differentiation and comparison with common-component loss gives equation~\eqref{eq:attentionparticipation}. \qed

\paragraph{Proof of Proposition~\ref{prop:welfare}.}
For alignment type $\nu_i$, local gross value is $q_i=v+\alpha\gamma_\Delta x+\alpha\nu_i\delta_\Delta r$. Investor surplus plus fee revenue equals $(q_i^2-f^2)/(2t)$. Reflection symmetry gives
\begin{equation}
\E[q_i^2]=q_0^2+\alpha^2\delta_\Delta^2r^2\sigma_{\nu,L}^2.
\end{equation}
For a contestable investor, choice surplus beyond the common component is
\begin{equation}
\E_\varepsilon\left|\alpha\nu_i\delta_\Delta r+\frac{\varepsilon}{2}\right|.
\end{equation}
Support-wide interiority implies $|\alpha\nu_i\delta_\Delta r|<h/2$. Uniform integration therefore gives
\begin{equation}
\E_\varepsilon\left|\alpha\nu_i\delta_\Delta r+\frac{\varepsilon}{2}\right|
=\frac{h}{4}+\frac{\alpha^2\nu_i^2\delta_\Delta^2r^2}{h}.
\end{equation}
Dropping constant $h/4$, integrating over $F_C$, cancelling fees against fund revenue, and subtracting both implementation costs yields equation~\eqref{eq:welfare}. Differentiation gives equations~\eqref{eq:Wx}--\eqref{eq:WDelta} and total differentiation gives equation~\eqref{eq:Wdecomp}. \qed

\paragraph{Proof of Proposition~\ref{prop:welfarethreshold}.}
Fee equilibrium implies $f^*=\mathcal N/\mathcal D$, $x^*=K_x\gamma_\Delta f^*$, $r^*=K_r\delta_\Delta f^*$, and $Q^*=2bf^*$. Substitution into equation~\eqref{eq:welfare} with $\eta=0$ yields equation~\eqref{eq:privatewelfare}. Since $\mathcal H'=b\alpha^2(\nu^2\phi_d-\phi_g)/\kappa<0$ and $\mathcal D'= -\ell\mathcal H'$, logarithmic fee derivative is $\rho_f=\ell\mathcal H'/\mathcal D<0$. Differentiating equation~\eqref{eq:privatewelfare}, using $(\gamma_\Delta^2)'=-1/2$, $(\delta_\Delta^2)'=1/2$, and $(\delta_\Delta^4)'=\delta_\Delta^2$, gives equation~\eqref{eq:Vprimeclosed}. Equation~\eqref{eq:Yclosed} and $\gamma_\Delta'/\gamma_\Delta=-1/[2(2-\Delta)]$ give equation~\eqref{eq:Yprimeclosed}. Finally,
\begin{equation}
\mathcal W^{*\prime}=V^{*\prime}+\eta Y^{*\prime}.
\end{equation}
Solving $\mathcal W^{*\prime}<0$ for $\eta$ gives equation~\eqref{eq:etathreshold} and the stated cases. \qed

\section{Extensions and robustness}
\label{sec:robustness}

\subsection{Smooth nonlinear ratings}
\label{sec:nonlinearrobustness}

Exact affinity inside a methodology cell keeps the core equilibrium transparent. Smooth nonlinear ratings produce nearby best responses when the affine game is strongly concave. Let $\widehat\pi_j(e_j;e_k)$ denote a fund's payoff under the affine rating map and $\pi_j(e_j;e_k)$ its payoff under a smooth map, where $e_j=(x_j,r_j)$. Suppose
\begin{equation}
\sup_{e_j\in\mathcal A_R}
|\pi_j(e_j;e_k)-\widehat\pi_j(e_j;e_k)|\leq\varepsilon_R
\label{eq:payoffperturb}
\end{equation}
for every rival action in compact feasible set $\mathcal A_R$, and suppose $\widehat\pi_j$ is $\mu$-strongly concave in own action.

\begin{lemma}[Best-response approximation]
\label{lem:perturbation}
Let $\widehat e_j(e_k)$ maximise $\widehat\pi_j(\cdot;e_k)$ and let $e_j^\circ(e_k)$ maximise $\pi_j(\cdot;e_k)$. Then
\begin{equation}
\|e_j^\circ(e_k)-\widehat e_j(e_k)\|
\leq2\sqrt{\frac{\varepsilon_R}{\mu}}
\label{eq:brbound}
\end{equation}
for every rival action. If rating remainders are second order and demand is Lipschitz, $\varepsilon_R=O(R^2)$.
\end{lemma}

Let $\widehat T$ and $T$ collect affine and nonlinear best responses. Equip joint actions with the maximum norm.

\begin{proposition}[Equilibrium approximation]
\label{prop:eqapprox}
Suppose $\widehat T$ is a contraction with modulus $\chi\in[0,1)$, both games have single-valued best responses on $\mathcal A_R^2$, and the payoff bound above holds. Let $\widehat e^*$ and $e^{\circ *}$ denote the affine and nonlinear equilibria. Then
\begin{equation}
\|e^{\circ *}-\widehat e^*\|_\infty
\leq
\frac{2}{1-\chi}
\sqrt{\frac{\varepsilon_R}{\mu}}.
\label{eq:eqbound}
\end{equation}
\end{proposition}

\paragraph{Proof of Lemma~\ref{lem:perturbation}.}
Strong concavity implies
\begin{equation}
\widehat\pi_j(\widehat e_j;e_k)-\widehat\pi_j(e_j^\circ;e_k)
\geq\frac{\mu}{2}\|e_j^\circ-\widehat e_j\|^2.
\end{equation}
Optimality of $e_j^\circ$ in the nonlinear problem and bound~\eqref{eq:payoffperturb} imply that the affine-payoff difference on the left-hand side does not exceed $2\varepsilon_R$. Rearrangement yields equation~\eqref{eq:brbound}. \qed

\paragraph{Proof of Proposition~\ref{prop:eqapprox}.}
Equilibrium fixed points satisfy $e^{\circ *}=T(e^{\circ *})$ and $\widehat e^*=\widehat T(\widehat e^*)$. Lemma~\ref{lem:perturbation} gives
\begin{align}
\|e^{\circ *}-\widehat e^*\|_\infty
&\leq\|T(e^{\circ *})-\widehat T(e^{\circ *})\|_\infty
+\|\widehat T(e^{\circ *})-\widehat T(\widehat e^*)\|_\infty\\
&\leq2\sqrt{\frac{\varepsilon_R}{\mu}}
+\chi\|e^{\circ *}-\widehat e^*\|_\infty.
\end{align}
Rearrangement yields equation~\eqref{eq:eqbound}. \qed

\subsection{Measurement uncertainty}
\label{sec:measurement}

Additive measurement errors can affect interior demand without changing methodology geometry. Evidence that disagreement coincides with lower trading activity and wider bid--ask spreads motivates the uncertainty channel \citep{quotb2026trading}. Funds choose efforts and fees before rating errors are realised. Investors observe published scores before allocating. Errors are independent of investor tastes and independent across funds; common and differential errors within a fund may have distinct variances. Measurement-uncertainty analysis maintains homogeneous attached alignment $\nu_j$ to isolate rating uncertainty from preference aggregation.

Let $\widetilde s_r=s_r+\varepsilon_r$ and define
\begin{equation}
\varepsilon_g=\frac{\varepsilon_1+\varepsilon_2}{2},
\qquad
\varepsilon_d=\frac{\varepsilon_1-\varepsilon_2}{2}.
\label{eq:errordecomp}
\end{equation}
Assume joint normality and $\Cov(\varepsilon_g,\varepsilon_d)=0$. Investor $i$ experiences rating-utility noise $\alpha(\varepsilon_g+\nu_i\varepsilon_d)$. Combining rating noise with normal reservation heterogeneity gives
\begin{equation}
P_j=\Phi\left(\frac{q_j}{\sigma_L(\nu_j)}\right),
\qquad
\sigma_L^2(\nu_j)=\sigma_x^2+\alpha^2(\sigma_g^2+\nu_j^2\sigma_d^2),
\label{eq:probitparticipation}
\end{equation}
where $q_j=v-f_j+\alpha\gamma_\Delta x_j+\alpha\nu_j\delta_\Delta r_j$. Balanced contestable investors have relative uncertainty $\sigma_C^2=\sigma_h^2+2\alpha^2\sigma_g^2$.

\begin{proposition}[Measurement precision and interior demand]
\label{prop:noise}
At a symmetric fixed-margin allocation, define
\begin{equation}
L_q=\frac{1-\lambda}{2}\frac{\phid(q/\sigma_L)}{\sigma_L},
\qquad
C_q=\lambda\frac{\phid(0)}{\sigma_C}.
\end{equation}
Interior first-order conditions are
\begin{equation}
\kappa x=m\alpha\gamma_\Delta(L_q+C_q),
\qquad
\kappa r=m\alpha\nu\delta_\Delta L_q.
\end{equation}
For $q>0$, greater local uncertainty lowers participation. Greater contestable uncertainty always weakens switching sensitivity. Local demand sensitivity falls when $|q|<\sigma_L$ and rises when $|q|>\sigma_L$:
\begin{align}
\frac{\partial P}{\partial\sigma_L}
&=-\frac{q}{\sigma_L^2}\phid(q/\sigma_L)<0,
\label{eq:uncertpart}\\
\frac{\partial C_q}{\partial\sigma_C}
&=-\lambda\frac{\phid(0)}{\sigma_C^2}<0,
\label{eq:uncertcontest}\\
\frac{\partial L_q}{\partial\sigma_L}
&=\frac{1-\lambda}{2}
\frac{\phid(q/\sigma_L)}{\sigma_L^2}
\left[\left(\frac{q}{\sigma_L}\right)^2-1\right].
\label{eq:uncertlocal}
\end{align}
\end{proposition}

\paragraph{Proof.}
Normal aggregation gives participation and contestable-choice probabilities. Differentiation supplies $L_q$, $C_q$, and equations~\eqref{eq:uncertpart}--\eqref{eq:uncertlocal}. Fixed-margin first-order conditions follow from marginal expected demand. \qed

Heterogeneous alignment replaces equation~\eqref{eq:probitparticipation} with an integral over $F_j(\nu)$. Resulting participation generally depends on the full alignment distribution because expected utility and measurement variance both vary with $\nu$.

\subsection{Timing, heterogeneous clienteles, and relative ratings}

Conflicting methodologies produce a simple timing result under fixed margins. Interior best responses depend on own clientele alignment but not on rival effort within a common coverage regime. Sequential and simultaneous positioning therefore coincide as long as no action induces a regime switch.

\begin{proposition}[Timing invariance under fixed margins]
\label{prop:timing}
Maintain Assumptions~\ref{ass:trust}--\ref{ass:contest} and any one coverage regime in Proposition~\ref{prop:regimes}. Provided both funds remain in the selected coverage regime, simultaneous and sequential positioning produce the same effort choices.
\end{proposition}

\paragraph{Proof.}
Partial-regime derivatives are $m\alpha b\gamma_\Delta-\kappa x_j$ and $m\alpha(\ell\nu_j+c\nu_C)\delta_\Delta-\kappa r_j$. Full-regime derivatives are $m\alpha c\gamma_\Delta-\kappa x_j$ and $m\alpha c\nu_C\delta_\Delta-\kappa r_j$. Rival effort enters profit only through an additive contestable-demand term. A follower therefore selects the same regime-specific maximiser after every admissible leader choice. Knowing the follower's regime-specific choice, the leader selects the same maximiser. Boundary first-order conditions inherit the same separability. \qed

Arbitrary alignment distributions enter partial-coverage demand through conditional means. For attached investors,
\begin{equation}
\int\frac{v-f+\alpha\gamma_\Delta x+\alpha\nu\delta_\Delta r}{t}\,dF_j(\nu)
=\frac{v-f+\alpha\gamma_\Delta x+\alpha\nu_j\delta_\Delta r}{t}.
\label{eq:heteromean}
\end{equation}
Independent uniform relative tastes give the same first-moment sufficiency for contestable investors. Type-specific saturation or correlation between relative taste and alignment makes higher moments relevant, but no perfect attachment-alignment correlation is required for Proposition~\ref{prop:direction}.

Fixed-bisector geometry is a canonical comparative-static path rather than a statement about every change in provider correlation. More generally, let $\gamma(\tau)>0$ fall and $\delta(\tau)>0$ rise while $u$ and $v$ remain fixed. Partial-coverage efforts satisfy
\begin{equation}
x_j(\tau)=\frac{m\alpha b\gamma(\tau)}{\kappa},
\qquad
r_j(\tau)=\frac{m\alpha\zeta_j\delta(\tau)}{\kappa}.
\end{equation}
Common effort then falls and absolute provider-specific effort rises without requiring the square-root normalisation. Rotating $u$ changes the holdings-outcome comparison through equation~\eqref{eq:rotatingoutcome}.

Rating providers may publish both absolute measures and relative labels. Let absolute score be $s_{rj}=b_r'z_j$ and displayed rating be
\begin{equation}
\widehat s_{rj}=H_r(s_{rj};s_{r,-j}),
\label{eq:relative}
\end{equation}
where $H_r$ maps absolute performance into a percentile or category. Portfolio costs depend on absolute scores because holdings must change to alter $s_{rj}$. Relative presentation changes demand but does not replace the underlying portfolio technology.

\subsection{Regime conditions}
\label{app:conditions}

Homogeneous-clientele partial coverage requires
\begin{equation}
0<v+\frac{m\alpha^2}{2\kappa}
\left[b(2-\Delta)+\nu^2\ell\Delta\right]<t.
\label{eq:partialcondition}
\end{equation}
For heterogeneous attached clienteles, both inequalities must hold after replacing $\nu$ by every element of the relevant support. Contestable interiority likewise requires equation~\eqref{eq:contestinterior} for every $\nu\in\operatorname{supp}(F_C)$.
Full coverage requires
\begin{equation}
v+\frac{m\alpha^2c(2-\Delta)}{2\kappa}\geq t.
\label{eq:fullcondition}
\end{equation}
Trust-region validity for fixed-margin partial effort requires
\begin{equation}
\left(\frac{m\alpha b\gamma_\Delta}{\kappa}\right)^2
+\left(\frac{m\alpha\nu\ell\delta_\Delta}{\kappa}\right)^2<R^2.
\label{eq:trustcondition}
\end{equation}
Endogenous fees additionally require \eqref{eq:feestability}, \eqref{eq:concavity},
\begin{equation}
0<v-f^*+\alpha\gamma_\Delta x^*
+\alpha\nu\delta_\Delta r^*<t,
\label{eq:feepartial}
\end{equation}
and $(x^*)^2+(r^*)^2<R^2$.

\section{Notation}
\label{app:notation}

\begin{longtable}{p{0.18\linewidth}p{0.72\linewidth}}
\toprule
Symbol & Definition \\
\midrule
\endfirsthead
\toprule
Symbol & Definition \\
\midrule
\endhead
$w_0,w$ & Financial benchmark portfolio and fund portfolio.\\
$N$ & Basis matrix for the budget-neutral portfolio tangent space.\\
$z$ & Reduced coordinates for the budget-neutral portfolio deviation $N z$.\\
$H$ & Positive-definite manager-level implementation-cost matrix.\\
$b_r$ & Reduced affine vector for rating methodology $r$.\\
$G$ & Methodology Gram matrix $BH^{-1}B'$.\\
$u,v$ & Fixed common and differential portfolio directions.\\
$\Delta$ & Systematic methodology disagreement.\\
$\gamma_\Delta,\delta_\Delta$ & Score loadings on common and differential effort.\\
$x$ & Common portfolio effort.\\
$r$ & Signed provider-specific portfolio effort.\\
$g,d$ & Common score component and signed score tilt.\\
$s^\perp$ & Provider-specific score component orthogonal to the common score direction.\\
$\kappa$ & Monetary weight on manager-level implementation cost.\\
$R$ & Radius of the feasible affine methodology cell.\\
$\lambda$ & Fraction of investors contestable across funds.\\
$\xi_i$ & Local investor $i$'s non-ESG mismatch.\\
$t$ & Local non-ESG differentiation.\\
$h$ & Dispersion of contestable relative tastes.\\
$\ell$ & Marginal local-demand coefficient, $(1-\lambda)/(2t)$.\\
$c$ & Marginal contestable-demand coefficient, $\lambda/(2h)$.\\
$b$ & Total common-effort demand coefficient, $\ell+c$.\\
$\phi_g,\phi_d$ & Fee-stage pass-through coefficients for common and provider-specific effort.\\
$v$ & Baseline investor surplus from fund participation.\\
$f_j$ & Fee charged by Fund $j$.\\
$\alpha$ & Willingness to pay for ESG attributes.\\
$p_i$ & Investor weight on Provider 1's methodology.\\
$\nu_i$ & Investor alignment, $2p_i-1$.\\
$\nu_j$ & Mean alignment in Fund $j$'s attached segment.\\
$\nu_C$ & Mean alignment among contestable investors.\\
$\sigma_{\nu,L}^2$ & Alignment variance within a symmetric attached clientele.\\
$\mu_{\nu,C}^{(2)}$ & Second moment of contestable investor alignment.\\
$\zeta_j$ & Demand-weighted effective alignment, $\ell\nu_j+c\nu_C$.\\
$J$ & Number of rating providers in the multiple-provider technology.\\
$m$ & Margin per investor in the fixed-margin benchmark.\\
$Q$ & Total participation in the two-fund market.\\
$\mathcal N,\mathcal H,\mathcal D$ & Closed-form fee and welfare auxiliaries defined in equations~\eqref{eq:welfareaux1}--\eqref{eq:welfareaux2}.\\
$\mathcal J_\nu$ & Aggregate matching coefficient from attached and contestable alignment dispersion.\\
$a$ & Loading vector for the holdings-based ESG outcome.\\
$y$ & Holdings-based ESG outcome $a'z$.\\
$\theta_u,\theta_v$ & Outcome loadings on common and differential portfolio directions.\\
$\sigma_g^2,\sigma_d^2$ & Variances of common and differential rating noise.\\
$\beta$ & Per-investor financial-performance loss coefficient.\\
$\eta$ & Social value assigned to holdings-based ESG exposure.\\
$\omega$ & Value of processing provider-specific methodology information.\\
$F_k$ & Conditional distribution of investor attention costs.\\
\bottomrule
\end{longtable}

\section*{Declarations}
\noindent\textbf{Conflict of interest.} No competing interests are reported.\\
\textbf{Funding.} No external funding supported the preparation of this manuscript.

\end{document}